\documentclass[12pt]{article}
\usepackage{epsf}

\makeatletter
\@addtoreset{equation}{section}

\makeatother

\newcommand\be{\begin{equation}}
\newcommand\ee{\end{equation}}
\newcommand\bea{\begin{eqnarray}}
\newcommand\eea{\end{eqnarray}}

\begin{document}

%\thispagestyle{empty}
%\begin{titlepage}
%
\begin{flushright}
UT-941\\
May, 2001 \\
\end{flushright}

\vspace*{3cm}

\begin{center}
{\Large\bf SUSY Breaking by Coexisting Walls}
\footnote{Parallel session talk at 8th International Symposium on 
Particles, Strings and Cosmology (PASCOS 2001), 
University of North Carolina at Chapel Hill, NC, USA, April 10-15, 2001.} \\
\vspace*{1cm}
{\large Nobuhito Maru} \\
\vspace*{5mm}
{\em Department of Physics, University of Tokyo, Tokyo 113-0033, JAPAN} \\

\vspace*{2cm}

\abstract{
Supersymmetry (SUSY) breaking without messenger fields is proposed. 
We assume that our world 
is on a wall and SUSY is broken only by the coexistence of another 
wall with some distance from our wall. 
The Nambu-Goldstone (NG) fermion is localized on the distant 
wall. 
Its overlap with the wave functions of physical fields on our wall gives 
the mass splitting of physical fields thanks to a low-energy theorem. 
We propose that this overlap provides a practical method to evaluate 
mass splitting in models with SUSY breaking due to the coexisting walls.} 
\end{center}

\newpage
%%%%%%%%%%%%%%%%%%%%%%%%%%%%%%%%%%%%%%%%%%%%%%%%%%%%%%%%%%%%%%%%%%%%%%
\section{Introduction}
%%%%%%%%%%%%%%%%%%%%%%%%%%%%%%%%%%%%%%%%%%%%%%%%%%%%%%%%%%%%%%%%%%%%%%
Recently, Brane World scenario \cite{ADD,RS} has opened new directions to 
the hierarchy problem, flavor physics, cosmology, astrophysics and so on. 
In this scenario, 
our world is considered as four dimensional topological objects 
embedded in higher dimensional spacetime.

On the other hand, 
it is well known that topological objects break SUSY partially 
in general. 
%\vspace*{-10mm}
%\begin{itemize}
%  \item 
Therefore, it is interesting to consider SUSY breaking 
in the context of the brane world scenario and 
also interesting to compare with other SUSY breaking mechanisms 
proposed so far, that is,  
%\end{itemize}
%\vspace*{-10mm}
gravity mediation, gauge mediation, anomaly mediation, gaugino mediation 
and radion mediation and so on.

In this talk, 
we propose a SUSY breaking mechanism due to the coexistence walls
\footnote{This talk is based on the work with N. Sakai, Y. Sakamura 
and R. Sugisaka \cite{MSSS}.}. 
An idea similar to ours has also been proposed and discussed 
in Ref. \cite{DS}.

%%%%%%%%%%%%%%%%%%%%%%%%%%%%%%%%%%%%%%%%%%%%%%%%%%%%%%%%%%%%%%%%%%%%%%
\section{SUSY breaking due to the other wall}
%%%%%%%%%%%%%%%%%%%%%%%%%%%%%%%%%%%%%%%%%%%%%%%%%%%%%%%%%%%%%%%%%%%%%%
In this section, our idea is briefly summarized. 
In order to avoid inessential complications, a toy model is discussed. 
Schematic picture is depicted in Fig 1. 
%
%%%%%%%%%%%%%%%%%%%%%%%%%%%%%%%%figure 1%%%%%%%%%%%%%%%%%%%%%%%%%%%%%%
\begin{figure}[h]
%\begin{center}
%\figurebox{22pc}{15pc}{} % to have a box alone
\epsfxsize=9cm
\epsfysize=6cm 
% will enlarge or reduce the postscript figures based on the xsize
\centerline{\epsfbox{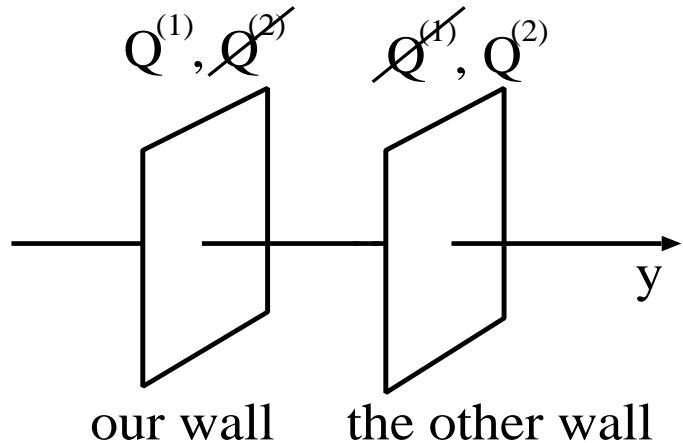}} % postscript image file name
\caption{Schematic picture of our setup.}
%\end{center}
\end{figure}
%%%%%%%%%%%%%%%%%%%%%%%%%%%%%%%%%%%%%%%%%%%%%%%%%%%%%%%%%%%%%%%%%%%%%%
%
The Bulk space-time is four dimensional. 
Two BPS domain walls are embedded. 
We assumed the bulk SUSY to be ${\cal N}=1$. 
We call one of the walls as ``our wall" where the matter field are localized 
and another wall as ``the other wall", which is the source of SUSY breaking. 
We have ${\cal N}=2$ SUSY in three dimension, which is denoted as 
$Q^{(1)}, Q^{(2)}$. 
$Q^{(1)}$ is conserved and $Q^{(2)}$ is spontaneously broken on our wall, 
and vice versa on the other wall. 
Remarkable features of this SUSY breaking mechanism are as follows. 
First, we need no SUSY breaking sector. 
Second, we also need no bulk messenger fields. 
Third, the half of SUSY is preserved on each wall 
{\em but completely broken in the whole system}. 

%%%%%%%%%%%%%%%%%%%%%%%%%%%%%%%%%%%%%%%%%%%%%%%%%%%%%%%%%%%%%%%%%%%%%%
\section{A Model with wall and antiwall}
%%%%%%%%%%%%%%%%%%%%%%%%%%%%%%%%%%%%%%%%%%%%%%%%%%%%%%%%%%%%%%%%%%%%%%
To illustrate the idea, we discuss a soluble example. 
The model is a minimal Wess-Zumino model in four dimensions. 
The Lagrangian is 
\begin{equation}
   {\cal L} = \Phi^{\dagger}\Phi|_{\theta^{2}\bar{\theta}^{2}}
   + W(\Phi)|_{\theta^{2}} + h.c.
\end{equation}
where
\begin{equation}
   W(\Phi)=\Lambda^{2}\Phi-\frac{g}{3}\Phi^{3}, 
\end{equation}
and $\Phi$ is a chiral superfield. 
One of the spatial direction $y$ is compactified on $S^{1}$ of 
the radius $R$.

This model has a nontrivial classical background solution \cite{STT}: 
\begin{equation}
A_{\rm cl}(y) = \frac{k\omega}{g}{\rm sn}(\omega(y-y_{0}),k), \quad
 \omega = \frac{\sqrt{2g}\Lambda}{\sqrt{1+k^{2}}}, \quad 
0 \le k \le 1. 
\end{equation}
By taking the radius $R$ appropriately, 
we obtain a wall-antiwall configuration. 
The properties of this configuration are the following. 
First, this configuration is static {\em but not stable}.
%\vspace*{-12mm}
Second, this instability is {\em not essential} 
to our SUSY breaking mechanism.
%\vspace*{-14mm}
Third, if walls are infinitely separated, 
this congifuration reduces to the familiar single wall solution, 
%
%\begin{eqnarray*}
$A_{{\rm cl}}(y) \to 
\frac{\Lambda}{\sqrt{g}}{\rm tanh}(\sqrt{g}\Lambda y)$.
%\end{eqnarray*}
%
Now, since we are interested in the effective theory on our wall, 
we would like to know the massless spectrum in the wall-antiwall background. 
There is a massless scalar field $\phi_{a,0}$, 
which implies a NG boson associated with 
the breaking of the translational invariance. 
The bosonic sector also has a tachyon $\phi_{a,-1}$, 
which implies the instabilitiy of the wall-antiwall system. 
On the other hand, 
there are two massless fermions 
$\varphi^{(1)}, \varphi^{(2)}$. 
One is a NG fermion associated with $Q^{(1)}$ SUSY breaking, 
another is one associated with $Q^{(2)}$ SUSY breaking.

\section{Estimation of $\Delta m^2$}
In this section, 
we estimate the mass splitting $\Delta m^2$ of the chiral multiplet 
localized on our wall. 
In order to do that, 
we obtain a three dimensional effective theory 
by integrating out massive modes.
In the effective theory, 
the effective Yukawa coupling 
$\sqrt{2}g_{\rm eff}a_{-1}\psi^{(1)}_{0}\psi^{(2)}_{0}$ is important 
because in our approach, 
the mass splitting $\Delta m^2$ is calculated through 
the supersymmetric analog of Goldberger-Treiman relation \cite{GT}: 
\begin{eqnarray}
g_{{\rm eff}} &=& -\frac{\Delta m^2}{\sqrt{2}f}, \quad 
\Delta m^2 \equiv m_B^2-m_F^2, \\
g_{\rm eff}&=& \int_{-\pi R}^{\pi R}{\rm d}y\phi_{a,-1}(y)
   \varphi^{(1)}_{0}(y)\varphi^{(2)}_{0}(y), 
\end{eqnarray}
where $f$ is the order parameter of SUSY breaking. 
Note that the effective Yukawa coupling $g_{\rm eff}$ becomes 
an overlap integral of modes localized on different walls. 
The result is displayed in Fig 2. 
%%%%%%%%%%%%%%%%%%%%%%%%%%%%%%%%%fig 2%%%%%%%%%%%%%%%%%%%%%%%%%%%%%%%%
\begin{figure}[h]
%\begin{center}
%\figurebox{22pc}{15pc}{} % to have a box alone
\epsfxsize=9cm
\epsfysize=6cm 
\centerline{\epsfbox{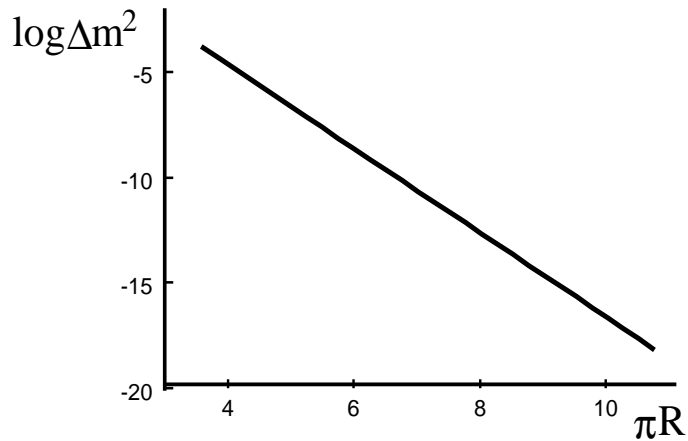}} % postscript image file name
\caption{The wall distance dependence of the mass splitting.}
%\end{center}
\end{figure}
%%%%%%%%%%%%%%%%%%%%%%%%%%%%%%%%%%%%%%%%%%%%%%%%%%%%%%%%%%%%%%%%%%%%%%
The mass splitting decays {\em exponentially} 
as the wall distance increases.
Note that this exponential suppression is obtained 
in spite of SUSY breaking at the classical level. 

\section{Matter fields}
We can also introduce a matter chiral superfield $\Phi_{\rm m}$ 
interacting with $\Phi$ through 
\begin{equation}
W_{\rm int}=-h\Phi\Phi_{\rm m}^{2}.
\end{equation}
Then, the matter fields is localized on our wall. 
When $h>g$, 
several light modes of $\Phi_{\rm m}$ are localized on our wall.

What is interesting here is that the mass splitting of the matter fields 
becomes larger for heavy fields\footnote{For details, see \cite{MSSS}}. 
This situation is easy to understand in our framework 
because the heavy modes have a large overlap with NG fermion 
localized on the other wall. 
We expect this phenomenon to be generic in our framework.

\section{The tachyonless model}
The wall-antiwall system discussed previously has a tachyonic mode, 
that is, the system is unstable. 
Here, we consider a tachyonless model to show such an instability 
does not necessarily appear. 

We consider the following model with two fields \cite{Gani}, 
%({\small Gani and Kudryavtsev ('99)})
\begin{equation}
{\cal L}=\Phi^{\dagger}\Phi|_{\theta^{2}\bar{\theta}^{2}}
   +X^{\dagger}X|_{\theta^{2}\bar{\theta}^{2}}
   +W(\Phi,X)|_{\theta^{2}}+h.c.,
\end{equation}
where
\begin{equation}
   W(\Phi,X)=\frac{m^{2}}{\lambda}\Phi-\frac{\lambda}{3}\Phi^{3}
   -\frac{\lambda}{4}\Phi X^{2}.
\end{equation}
This model has four supersymmetric vacua 
$X=0, \Phi = \pm m/\lambda$ and $\Phi=0, X = \pm 2m/\lambda$. 
{\em Tachyonless Non-BPS} configuration is constructed 
from a superposition of BPS walls: 
\begin{eqnarray}
\phi_{\rm cl}(y) &=& \frac{m}{2\lambda} 
\left( \tanh\frac{m y}{2}
 -\tanh\frac{m(y-d)}{2} \right), \\
\chi_{\rm cl}(y) &=& \frac{\sqrt{2}m}{\lambda} 
\left( \sqrt{1-\tanh\frac{m y}{2}}
-\sqrt{1+\tanh\frac{m(y-d)}{2}} \right). 
\end{eqnarray}
Since the vacua at $y=-\infty$ and $y=\infty$ are {\em different}, 
this configuration is {\em stable}. 
We have checked that 
the mass splitting can also be calculated by the overlap integral 
and exponentially suppressed in this model.

\section{Summary}
\begin{itemize}
 \item We have investigated a mechanism of SUSY breaking
 due to the {\em coexistence} of walls. 
 \item The mechanism does {\em not} need {\em any SUSY breaking sector} 
 or {\em messenger bulk fields}. 
 \item The effective SUSY breaking scale observed on our wall becomes
 {\em exponentially small} as the distance between two walls grows. 
 \item SUSY breaking effects we observe can be calculated from
 the {\em overlap} of mode functions. 
 \item The mass splittings of the matter fields becomes
 large for heavier modes. 
 \item The {\em instability} of the configuration in the model 
 discussed here is {\em not essential} to our SUSY breaking mechanism. 
\end{itemize}

\begin{center}
{\bf Acknowledgments}
\end{center}
The author would like to thank N. Sakai, Y. Sakamura and  R. Sugisaka for 
collaboration. 
He would like to thank the organizers of PASCOS 2001 
for giving me the opportunity of the talk. 
This work was supported by the Japan Society for the Promotion of Science 
for Young Scientists (No.08557).

%\section*{Appendix}

\end{document}